\documentclass{aa}  

\usepackage{graphicx}
\usepackage{txfonts}
\usepackage{epstopdf}
\usepackage{textcomp}

\def\ltapprox{\raise 2pt \hbox {$<$} \kern-1.1em \lower 5pt \hbox {$\approx$}}
\def\ltsim{\raise 2pt \hbox {$<$} \kern-1.1em \lower 4pt \hbox {$\sim$}}
\def\gtsim{\raise 2pt \hbox {$>$} \kern-1.1em \lower 4pt \hbox {$\sim$}}

\def\ie{{\it i.e.,~}}
\def\eg{{\it e.g.,~}}
\def\gtsim{\; \raise0.3ex\hbox{$>$\kern-0.75em \raise-1.1ex\hbox{$\sim$}}\; }
\def\ltsim{\; \raise0.3ex\hbox{$<$\kern-0.75em \raise-1.1ex\hbox{$\sim$}}\; }

\begin{document}

   \title{On the occurrence of Radio Halos in galaxy clusters}

   \subtitle{Insight from a mass-selected sample}

   \author{V. Cuciti
          \inst{1,2}
          \and
          R. Cassano\inst{2} 
          \and
          G. Brunetti\inst{2}
          \and
          D. Dallacasa\inst{1,2}
          \and
          R.Kale\inst{3}
          \and
          S. Ettori\inst{4,5}
          \and
          T. Venturi\inst{2}
          }

   \institute{Dipartimento di Fisica e Astronomia, Universit\`{a} di Bologna, via Ranzani 1, 40127 Bologna, Italy\\
              \email{vcuciti@ira.inaf.it}
         \and
             INAF-Istituto di Radioastronomia, via P. Gobetti 101, 40129 Bologna, Italy
             \and
             National Centre of Radio Astrophysics, TIFR, Pune, India
             \and INAF/Osservatorio Astronomico di Bologna, via Ranzani 1, I-40127 Bologna, Italy
             \and INFN, Sezione di Bologna, viale Berti Pichat 6/2, I-40127, Bologna, Italy}

   \date{Received ; accepted }

 
  \abstract
   {Giant radio halos (RH) are diffuse Mpc-scale synchrotron sources detected in a fraction of massive and merging galaxy clusters. An unbiased study of the statistical properties of RHs is crucial to constrain their origin and evolution.}
   {We aim at investigating the occurrence of RHs and its dependence on the cluster mass in a SZ-selected sample of galaxy clusters, which is as close as possible to be a mass-selected sample. Moreover, we analyse the connection between RHs and merging clusters.}
   {We select from the Planck SZ catalogue (Planck Collaboration XXIX 2014) clusters with $M\geq 6\times 10^{14} M_\odot$ at $z=0.08-0.33$ and we search for the presence of RHs using the NVSS for $z<0.2$ and the GMRT RH survey (GRHS, Venturi et al. 2007, 2008) and its extension (EGRHS, Kale et al. 2013, 2015) for $0.2<z<0.33$. We use archival \textit{Chandra} X-ray data to derive information on the clusters dynamical status.}
   {We confirm that RH clusters are merging systems while the majority of clusters without RH are relaxed, thus supporting the idea that mergers play a fundamental role in the generation of RHs. We find evidence for an increase of the fraction of clusters with RHs with the cluster mass and this is in line with expectations derived on the basis of the turbulence re-acceleration scenario. Finally, we discuss the effect of the incompleteness of our sample on this result.}
   {}


   \maketitle
%

\section{Introduction}

Clusters of galaxies are the largest and most massive bound systems in the Universe. They form and grow at the intersection of cosmic filaments where matter and galaxies get together as a consequence of the 
gravitational collapse. Mergers between clusters of galaxies are among the most energetic events in the Universe as they release
energies of $\sim 10^{63}-10^{64}$ erg in few Gyrs. Although most of this energy is dissipated to heat the intra cluster medium (ICM) up to temperature of $\sim10^7-10^8\,^{\circ}K$, part of this energy is channelled into the acceleration of relativistic particles and amplification of magnetic fields in the ICM (see \eg Brunetti \& Jones 2014, for a review). Diffuse Mpc-scale synchrotron radio emission observed in a growing number of galaxy clusters is the most direct and compelling evidence of this activity.
Non-thermal radio emission from galaxy clusters is observed in the form of giant radio halos (RH), located at the cluster center with morphology similar to that of the X-ray emission, and radio relics, located at the cluster outskirts and characterised by elongated shapes (\eg Feretti et al. 2012). The emerging theoretical picture is that radio relics trace shock waves propagating out of the cluster cores, whereas radio halos trace turbulent regions in clusters, where particles are trapped and re-accelerated during mergers (\eg Brunetti \& Jones 2014). The comparison between thermal and non-thermal properties of galaxy clusters provides important information on the complex mechanisms that generate the observed radio emission.
According to models based on turbulent acceleration, the formation history of RHs depends on the cluster merging rate throughout cosmic epochs and on the mass of the hosting clusters, which ultimately sets the energy budget available for the acceleration of relativistic particles. 
In their simplest form, these models predict a steepening in the spectra of RHs at a frequency $\nu_s$ which directly depends on the energetics of the merger (\textit{i.e.} on the cluster mass). Therefore a key expectation is that typical RHs should preferentially be found in massive objects undergoing energetic merging events, whereas  they should be rarer in less massive merging-systems and absent in relaxed clusters (\eg Cassano \& Brunetti 2005).
Smaller systems undergoing less energetic merging events are expected to produce RHs with increasingly steep spectra (lower $\nu_s$) which become underluminous at higher frequencies.
This implies the existence of RHs with ultra-steep radio spectra (USSRH, $\alpha>1.5$, with $f(\nu)\propto \nu^{-\alpha}$) that should become better visible at low radio frequency (Cassano et al. 2006; Brunetti et al. 2008, Dallacasa et al. 2009).

A first statistical measurement of the occurrence of giant RHs in galaxy clusters has been obtained through the ``GMRT RH Survey'' (Venturi et al. 2007, 2008; GRHS hereafter) and its extension (the EGRHS, Kale et al. 2013, 2015). This survey is restricted to clusters in the redshift range $0.2-0.4$. It confirmed that RHs are hosted in only $\sim20-30\%$ of X-ray luminous ($L_X(0.1-2.4\,\mathrm{keV})\geq5\times10^{44}$ erg/s) clusters and found that clusters branch into two populations: RHs trace a correlation between $P_{1.4}$ and $L_X$, whereas radio-undetected clusters (upper limits) lie about 1 order of magnitude below the correlation (\eg Brunetti et al. 2007). Importantly, this bimodal split can be traced to clusters dynamics: RHs are always associated to merging systems\footnote{A possible outlier is the RH recently discovered in the cool-core cluster CL1821+643 (Bonafede et al. 2014).} while clusters without RH are typically relaxed (\eg Cassano et al. 2010). 

The recent advent of cluster surveys via the SZ effect (\ie with the {\it Planck} satellite) has enabled the construction of unbiased cluster samples that are almost mass-selected, due to the tight relation between the total SZ signal, $Y_{500}$, when integrated within $R_{500}\footnote{$R_{500}$ is the radius corresponding to a total density contrast 500$\rho_c(z)$, $\rho_c(z)$ is the critical density.}$, and the cluster mass, $M_{500}$ (Motl et al. 2005; Nagai 2006).
The fraction of clusters with RHs appears larger in SZ-selected cluster samples with respect to that derived from X-ray samples (Sommer \& Basu 2014). 
Earlier studies were unable to observe a bimodal behaviour of clusters with RH and radio-undetected systems in the radio-SZ properties (Basu 2012). However, thanks to the improved statistics, more recently Cassano et al. (2013) demonstrated the presence of a  bimodal split also in the radio-SZ diagram, for $Y_{500} > 6\times 10^{-5}$ Mpc$^2$, and confirmed that such split is tightly connected with the dynamical properties of the hosting clusters. This result provides strong evidence, complementary to X-ray studies, that mergers play a key role in the formation of RHs. However, the relatively low mass completeness ($\sim50\%$) of the Cassano et al. (2013) sample did not allow to measure the occurrence of RHs and in particular to study such occurrence as a function of the cluster mass. 

In order to provide an unbiased measure of the fraction of clusters hosting RHs and of its dependence on the cluster mass, we selected from the PSZ catalogue clusters with $M\gtsim 6\times10^{14}\,M_{\odot}$ in the redshift range $z\simeq 0.08-0.33$. In this way we obtained a sample of 75 clusters with mass completeness $> 80\%$, 57 of them have available radio information. Here we report on the statistical analysis of these 57 clusters that constitute a sample with mass completeness $\sim 63 \%$ (see Sect.~\ref{sec:sample} and \ref{sec:completeness}). The addition of the remaining 18 clusters without radio information will allow to achieve a completeness in mass $>80\%$. Deep JVLA and GMRT observations of these 18 clusters are in progress and results will be presented in a follow-up paper.

\noindent We also used the {\it Chandra} X-ray data, available for most of the clusters in the sample, to investigate their dynamical status and the connection with the radio properties.

In Sect.~2 we describe the selection of the cluster sample; in Sect.~3 we report on the analysis of NVSS data of low-z clusters; in Sect.~4 we derive the cluster dynamical status. In Sect.~5 we derive the occurrence of clusters with giant radio halos and in Sect.~6 we investigate the RH-merger connection. In Sect.~7 we discuss the effect of the sample completeness on the results and we report our conclusions in Sect.~8.


\section{Cluster sample selection}
\label{sec:sample}
 \begin{figure}
   \centering
  \includegraphics[width=7cm]{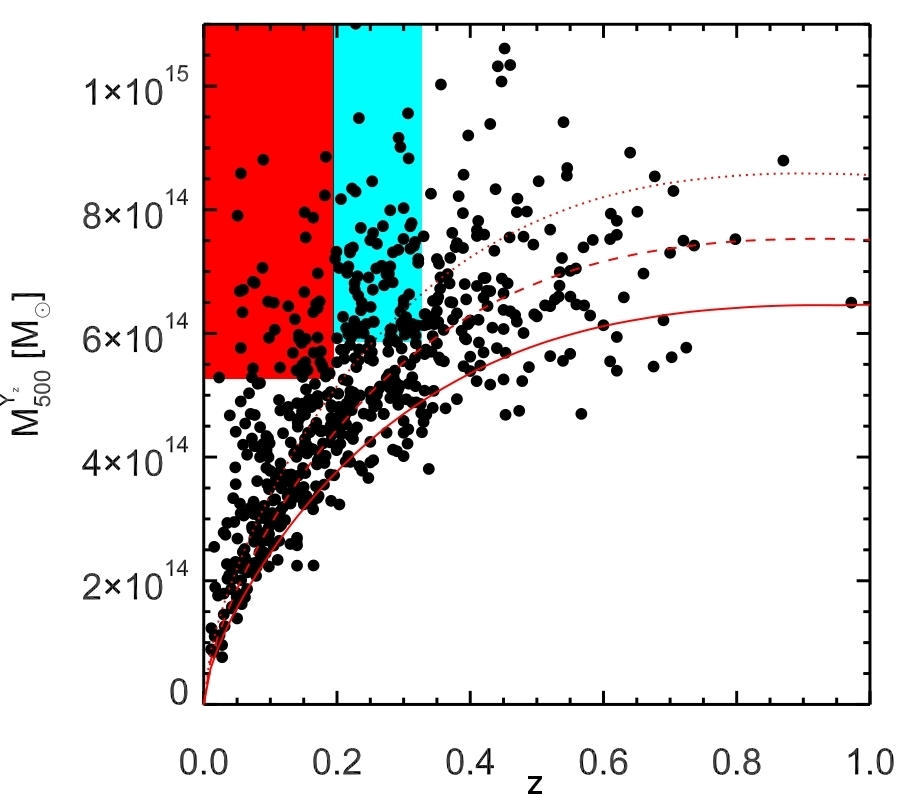}
   \caption{\small {Average mass limit computed from the average noise over the sky for the PSZ catalogue. The dotted, dashed and solid lines show the Planck mass limit at 80, 50 and 20 \% completeness respectively. The rectangles show the regions from where we extracted our sample: the red rectangle for the low redshift sample, the cyan retangle for the high redshift one. Adapted from Planck Collaboration XXIX 2014.}}\label{fig:planck}
   \end{figure}

We used the Planck SZ cluster catalogue (PSZ, Planck Collaboration XXIX 2014) to select a sample of massive galaxy clusters. This catalogue consists of 1227 objects derived from SZ effect detections using the first 15.5 months of Planck satellite observations. It contains 861 confirmed clusters and 366 cluster candidates. 
To date the Planck sample is the largest SZ-selected cluster sample (six times the size of the Planck Early SZ, Planck Collaboration 2011) and  the deepest all-sky catalogue. It spans the broadest cluster mass range from 0.1 to $1.6\times 10^{15} \,M_\odot$, with redshift up to about one.

From the PSZ catalogue we selected clusters with $M_{500}\gtrsim6\times10^{14} \,M_\odot$\footnote{The values of $M_{500}$ in the PSZ catalogue are obtained from $Y_{500}$ as described in Sect.7.2.2 in Planck Collaboration XXIX 2014.} and redshift $0.08<z<0.33$. To maximize the radio coverage we adopted a declination limit $\delta>-31$\textdegree and $|b|\geq 20$\textdegree ($|b|$ is the galactic latitude) for clusters at $z>0.2$, that coincides with that of the GMRT radio surveys. On the other hand at lower redshift  we adopted $\delta>-40$\textdegree ~to ensure a follow up from the NVSS radio survey (Condon et al. 1998).


\noindent
Among the 54 clusters at $z>0.2$, 34 belong to the EGRHS and thus have deep radio observations, additional 2 clusters have literature information (namely PSZ1 G205.07-62.94 and PSZ1 G171.96-40.64).

\noindent
For targets in the redshift range $0.08-0.2$ we collected data from the literature (14 clusters) and analysed data from the NVSS radio survey (Condon et al. 1998) for the remaining 7 clusters (see Sect.~\ref{Sect.NVSS}).

In Fig.\ref{fig:planck} we show the $M_{500}-z$ distribution of the Planck clusters detected over 83.7\% of the sky, together with the Planck mass limit corresponding to the 80, 50 and 20\% completeness of the catalogue (Planck Collaboration XXIX 2014). The red and blue boxes show the regions from where we selected our cluster sample: the low-z sample and the high-z sample have a mass-completeness of $\sim$90\% and 80\%, respectively.
\noindent
The sample with radio information consists of 57 clusters (21 at $z=0.08-0.2$ and 36 at $z=0.2-0.33$) with a completeness in mass of $\sim 90\%$ at low redshift and $\sim 53\%$ ($0.8\times \frac{36}{54}=0.53$) at higher redshift. The 57 clusters and their properties are listed in Tab.~\ref{tab:completesample}.  

 \begin{table*}
 
 \center
\caption{Total sample clusters properties}             
\label{tab:completesample}      
\begin{tabular}{l c c c c c c }     
\hline      
\hline
cluster name &    RA    &      Dec&         z& $M_{500}$ &  Radio info& X-ray info\\
& & & & $(10^{14}\,M_\odot$)& & \\
\hline
A1437&      12 00 22.3&    $+$03 20 33.9&    0.134&       5.69& no RH$^{\Large *}$& M $\surd$ \\
A2345&      21 27 06.8&    $-$12 07 56.0&    0.176&       5.71& Relics$^4$& M $\surd$\\
A2104&      15 40 08.2&    $-$03 18 23.0&    0.153&       5.91& no RH$^{\Large *}$& M $\surd$ \\
Zwcl 2120.1+2256& 21 22 27.1& $+$23 11 50.3& 0.143&       5.91&no RH$^{\Large *}$ & M $\surd$\\
RXC J0616.3-2156& 06 16 22.8& $-$21 56 43.4& 0.171&       5.93& no RH$^{\Large *}$& M $\surd$\\
A1413&      11 55 18.9&     $+$23 24 31.0&   0.143&       5.98& MH $^5$& R $\surd$\\  
A1576&      12 37 59.0&     $+$63 11 26.0&     0.302&       5.98& UL$^6$& R$^{26}$  \\
A2697&      00 03 11.8&     $-$06 05 10.0&     0.232&       6.01& UL$^2$& R $^{x}$\\
Z5247&	12 33 56.1&		$+$09 50 28.0&	0.229&		6.04& RH$^7$ & M $\surd$\\
Zwcl 0104.9+5350& 01 07 54.0& $+$54 06 00.0& 0.107&       6.06& RH$^8$& --\\
RXC J0142.0+2131& 01 42 02.6& $+$21 31 19.0&   0.280&       6.07& UL$^6$& R $^{26}$\\
A1423&      11 57 22.5&     $+$33 39 18.0&     0.214&       6.09& UL$^2$& R$^{25}$\\
RXC J1314.4-2515& 13 14 28.0& $-$25 15 41.0&   0.244&       6.15& RH$^1$& M$^{x}$\\
A2537&      23 08 23.2&     $-$02 11 31.0&     0.297&       6.17& UL$^2$& R$^{25}$\\  
A68&		00 37 05.3&		$+$09 09 11.0&	0.255&		6.19& UL$^7$ & M $\surd$\\
A1682&      13 06 49.7&     $+$46 32 59.0&     0.226&       6.20& RH$^2$& M$^{25}$\\
A1132&      10 58 19.6&     $+$56 46 56.0&	  0.134&       6.23& no RH$^3$& M $\surd$ \\
RXJ1720.1+2638&  17 20 10.1& $+$26 37 29.5&  0.164&       6.34& MH$^9$ & R $\surd$\\
A781&       09 20 23.2&     $+$30 26 15.0&	  0.295&       6.36& UL$^2$& M$^{25}$\\
A2218&      16 35 51.6&     $+$66 12 39.0&   0.171&       6.41& RH$^3$& M$^{28}$ $\surd$\\
A3411&      08 41 55.6&     $-$17 29 35.7&   0.169&       6.48& RH$^{10}$& M $\surd$\\
Zwcl 0634.1+4750& 06 38 02.5& $+$47 47 23.8& 0.174&       6.52& suspect$^{\Large *}$ & M $\surd$\\
A3888&      22 34 26.8&     $-$37 44 19.1&   0.151&       6.67& suspect$^{\Large *}$& M?$^{x, 29}$  \\
A3088&      03 07 04.1&     $-$28 40 14.0&     0.254&       6.71& UL$^2$& R$^{26}$\\
A2667&      23 51 40.7&     $-$26 05 01.0&     0.226&       6.81& UL$^2$& R$^{25}$\\
A521&       04 54 09.1&     $-$10 14 19.0&     0.248&       6.91& RH$^{11,US}$& M$^{25}$\\
A2631&      23 37 40.6&     $+$00 16 36.0&     0.278&       6.97& UL$^2$& M$^{25}$\\
A1914&      14 26 03.0&     $+$37 49 32.0&   0.171&       6.97& RH$^{12}$& M $\surd$\\
RXC J1504.1-0248& 15 04 07.7&  $-$02 48 18.0&  0.215&       6.98& MH$^{13}$& R$^{25}$\\
A520&       04 54 19.0&     $+$02 56 49.0&     0.203&       7.06& RH$^{14}$& M$^{25}$\\
A478&       04 13 20.7&     $+$10 28 35.0&   0.088&       7.06& MH$^{15}$& R $\surd$\\
A773&       09 17 59.4&     $+$51 42 23.0&     0.217&       7.08& RH$^{14}$& M$^{25}$\\
A1351&      11 42 30.8&     $+$58 32 20.0&     0.322&       7.14& RH$^{16}$& M $\surd$\\
A115&       00 55 59.5&     $+$26 19 14.0&   0.197&       7.21& Relic$^{14}$& M $\surd$\\
A1451&      12 03 16.2&     $-$21 32 12.7&   0.199&       7.32& suspect$^{\Large *}$& M $^{x}$ \\
PSZ1 G205.07-62.94& 02 46 27.5& $-$20 32 5.29& 0.310& 7.37&      no RH$^p$&M$^{x}$\\ 
A2261&      17 22 17.1&     $+$32 08 02.0&      0.224&       7.39& UL$^6$& R$^{25}$\\
RXCJ2003.5-2323& 20 03 30.4& $-$23 23 05.0&    0.317&       7.48& RH$^1$& M$^{25}$\\
A2552&	23 11 26.9&		$+$03 35 19.0&	0.300&		7.53& RH?$^7$ & R? $\surd$\\
A3444&      10 23 50.8&     $-$27 15 31.0&     0.254&       7.62&  MH$^7$& R $\surd$\\
S780&       14 59 29.3&     $-$18 11 13.0&     0.236&       7.71& MH$^7$& R$^{25}$\\
A2204&      16 32 45.7&     $+$05 34 43.0&   0.151&       7.96& MH$^{15}$& R $\surd$\\
A1758a&     13 32 32.1&     $+$50 30 37.0&     0.280&       7.99& RH$^{17}$& M$^{25}$\\ 
A209&	    01 31 53.0&     $-$13 36 34.0&     0.206&	   8.17& RH$^1$& M$^{25}$\\
A665&       08 30 45.2&     $+$65 52 55.0&    0.182&       8.23& RH$^3$& M $\surd$\\
A1763&		13 35 17.2&	    $+$40 59 58.0&	  0.228&	  	   8.29& no RH$^2$& M $\surd$\\
RXC J1514.9-1523& 15 14 58.0& $-$15 23 10.0&   0.223&	   8.34& RH$^{18,c}$& M $\surd$\\
A1835&		14 01 02.3&		$+$02  52 48.0&    0.253&		   8.46& MH$^{19}$& R $\surd$\\
A2142&      15 58 16.1&     $+$27 13 29.0&   0.089&       8.81& RH$^{24}$& M$^{30}$ $\surd$\\
A1689&      13 11 29.5&     $-$01 20 17.0&   0.183&       8.86& RH$^{20}$& M$^{27}$ $\surd$\\ 
A1300&		11 31 56.3&		$-$19 55 37.0&	  0.308&		   8.83& RH$^{21,c}$& M$^{25}$\\
A2390&		21 53 34.6&		$+$17 40 11.0&    0.234&		   9.48& MH$^{12}$& R$^{25}$\\
A2744&		00 14 18.8&		$-$30 23 00.0&	  0.307&		   9.56& RH$^{14}$& M$^{25}$\\
A2219&		16 40 21.1&		$+$46 41 16.0&	  0.228&		   11.01& RH$^{12}$& M$^{25}$\\
PSZ1 G171.96-40.64& 03 12 57.4& $+$08 22 10& 0.270&	   11.13& RH$^{22,c}$& M$^x$\\
A697&		08 42 53.3&		$+$36 20 12.0&    0.282&		   11.48& RH$^{32,US}$& M$^{25,31}$\\
A2163&		16 15 46.9&		$-$06 08 45.0&	  0.203&		   16.44& RH$^{23}$& M$^{25}$\\
\hline  
\end{tabular} 
\begin{flushleft}
RH = Radio Halo, MH = Mini-Halo, UL = Upper Limit, M=merger, R= relaxed.
$^1$ Venturi et al. (2007), $^2$ Venturi et al. (2008), $^3$ Giovannini \& Feretti (2000), $^4$ Bonafede et al. (2009), $^5$ Govoni et al. (2009), $^6$ Kale et al. (2013), $^7$ Kale et al. (2015) , $^8$ van Weeren et al. (2011), $^9$Giacintucci et al. (2014b), $^{10}$van Weeren et al. (2013), $^{11}$Brunetti et al. (2008), $^{12}$Bacchi et al. (2003), $^{13}$Giacintucci et al. (2011a), $^{14}$Govoni et al. (2001), $^{15}$ Giacintucci et al. (2014a), $^{16}$ Giacintucci et al. (2009), $^{17}$Giovannini et al. (2006), $^{18}$Giacintucci et al. (2011b), $^{19}$Murgia et al. (2009), $^{20}$Vacca et al. (2011), $^{21}$Venturi et al., 2013, $^{22}$Giacintucci et al. (2013), $^{23}$Feretti et al. (2001), $^{24}$Farnsworth et al. (2013), $^{25}$Cassano et al. (2010), $^{26}$Cassano et al. (2013), $^{27}$Andersson \& Madejski (2004), $^{28}$Pratt et al. (2005), $^{29}$Wei\ss man et al. (2013a), $^{30}$Owers et al. (2011), $^{31}$Girardi et al. (2006), $^{32}$Macario et al., 2010, $^{US}$ Ultra Steep Spectrum RH, $^c$ candidate USSRH,  $^p$Ferrari et al. (private communication), $^{x}$ visual inspection of XMM-Newton image, $^{\Large *}$NVSS data analysed in this paper, $\surd$ X-ray Chandra data analysed in this paper. 
\end{flushleft}  
\end{table*}

\section{The low-z sample and the NVSS data analysis}
\label{Sect.NVSS}

We use the NRAO VLA Sky Survey (NVSS, Condon et al. 1998) to investigate the presence of cluster-scale diffuse emission in the 7 clusters of the low-z sample missing radio information in the literature. The NVSS is a radio survey performed at 1.4 GHz with the Very Large Array (VLA) in D and DnC configuration. It covers the sky north of $\delta=-40$\textdegree, it has an angular resolution of 45'' and a surface brightness rms of $\sim0.45$ mJy/beam. 

The low-z sample includes clusters with $0.08<z<0.2$ and $M_{500}\gtrsim 5.7\times10^{14} \,M_\odot$. Radio interferometers suffer from the lack of sampling at short baselines, resulting in decreased sensitivity to emission on large spatial scales. For this reason we adopted a lower redshift limit of $z>0.08$. Indeed, Farnsworth et al. (2013) showed that on scales $\gtrsim11$ arcmin, that correspond to a 1 Mpc halo at $z\sim 0.08$, less than 50\% of the total flux density is recovered with a NVSS snapshot observation.

The upper redshift limit ($z<0.2$) and the minimum mass are set by the angular resolution and sensitivity of the NVSS.
The NVSS beam of $45''$ corresponds to  $\sim150$ kpc at $z=0.2$ and this does not allow to separate discrete sources from residual diffuse emission at higher redshift.
 
\noindent
From Eq. 9 in Cassano et al. (2012), adopting the sensitivity and resolution of the NVSS, we derived the minimum $P_{1.4}$ of a detectable RH. The minimum mass $M_{500}=5.7\times10^{14}M_\odot$ has been then derived assuming the $P_{1.4}-M_{500}$ correlation (Eq. 14 in Cassano et al. 2013).

With these selection criteria, the low-z sample is made of 21 clusters. For 14 of these clusters we found information in the literature on the presence/absence of cluster-scale radio emission that are based on pointed VLA/WSRT observations.

\subsection{NVSS data analysis}
\label{Sec:NVSS_analysis}
Here we describe the NVSS data analysis carried out to investigate the presence of diffuse radio emission in the 7 clusters (marked with * in Tab.~\ref{tab:completesample}) which lack literature radio information. To improve the quality of the radio images, \ie to lower the rms noise and  reduce the contribution of noise pattern, we reprocessed the NVSS fields of these 7 clusters. Data were analysed using the NRAO Astronomical Image Processing System (AIPS). We calibrated the NVSS dataset and we obtained the images of the pointings containing the cluster, then we combined them with the task FLATN. This procedure, known as the mosaic technique, is fundamental especially when the cluster falls at the border of the primary beam, because the signal to noise ratio decreases with the distance from the pointing position. For the 7 reprocessed clusters we reached an average rms$\approx0.25$ mJy/beam, which is $\sim2$ times better than the nominal NVSS noise. 

\begin{figure*}
\centering
\includegraphics[scale=0.32]{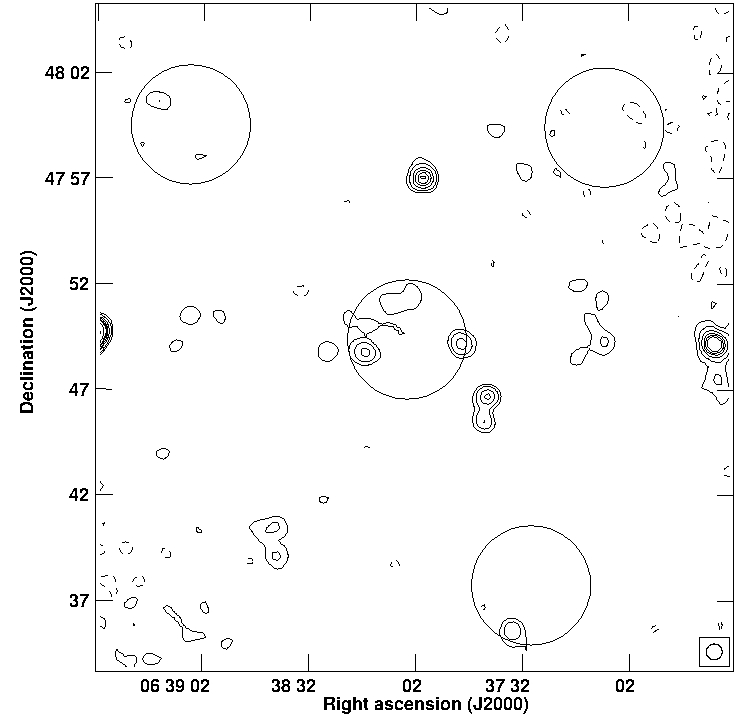}
\includegraphics[scale=0.32]{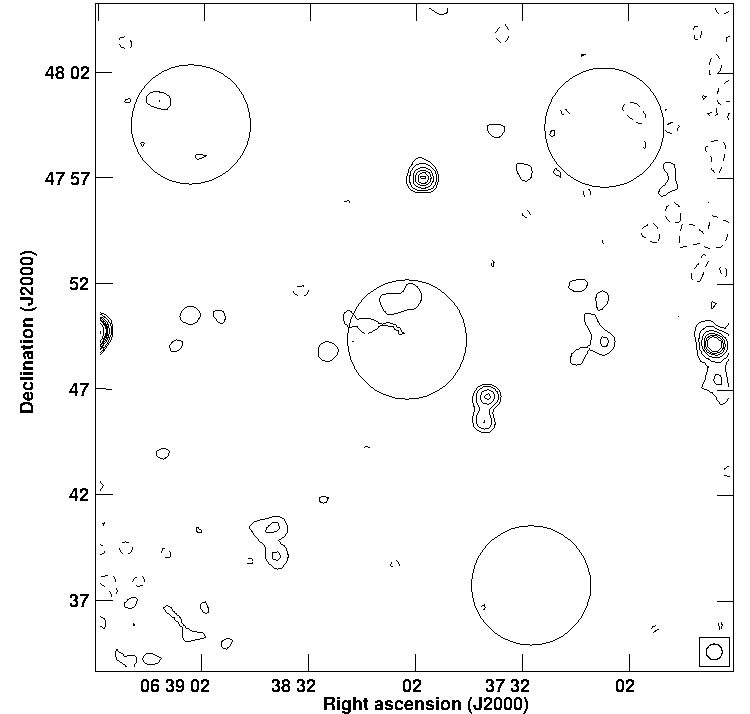}
\caption{\small{Zwcl0634.1+4750 NVSS map, the contour levels are $0.66\times (-1, 1, 2, 4, 6, 8, 10, 16, 32, 64)$ mJy b$^{-1}$. The 1$\sigma$ level is 0.22 mJy b$^{-1}$. In both panels the region where we extracted the flux densities are shown (solid circles). The central region has a diameter of 1 Mpc and is centred on the centroid of the X-ray emission; the other 3 areas are the so-called control fields. With the task BLANK we masked the discrete sources in the central region and the one that falls in the lower control field (left panel).}}\label{fig:A3888}
\end{figure*}

None of these clusters show clear diffuse cluster-scale radio emission, however, we further investigated the possible presence of residual emission in the central regions of these clusters. Specifically, we selected on each map a 1 Mpc sized circle centred on the centroid of the cluster X-ray emission. With the task BLANK, we masked the discrete sources in the cluster that show contours at least at the 6$\sigma$ level, then we measured the residual diffuse flux density (RDF, hereafter) in the circular region. We compared the RDF with the flux densities measured in other areas of the same size taken around the cluster (3 for each cluster), \ie ``control fields''. In order to make a consistent comparison we normalized both the RDF and the control field flux densities (CFF, hereafter) to the number of pixels enclosed in a circle of 1 Mpc diameter after masking the discrete sources.
An example of this procedure is reported in Fig.~\ref{fig:A3888}, applied to the case of Zwcl0634.1+4750. We stress that few NVSS beams correspond to 100-300 kpc. Consequently the use of the task BLANK in the case of relatively bright central sources is expected to remove also diffuse emission on these scales. This is particularly problematic for the case of mini-halos that however are not the central focus of this paper.


To test the reliability of this procedure, we also applied it to 3 known RH clusters (A3411, A2218, Zwcl 0104.9+5350) and to the mini-halo in RXJ1720.1+2638 that belong to our sample, but have information from the literature. 

\begin{figure*}
\centering
\includegraphics[scale=0.4]{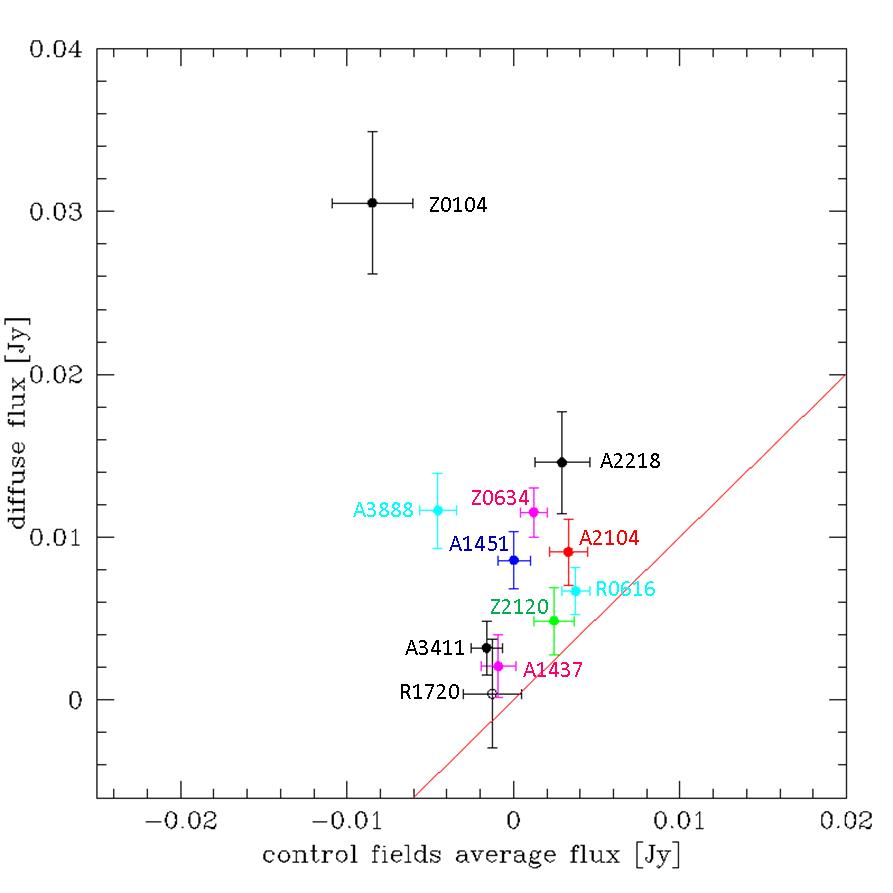}
\includegraphics[scale=0.4]{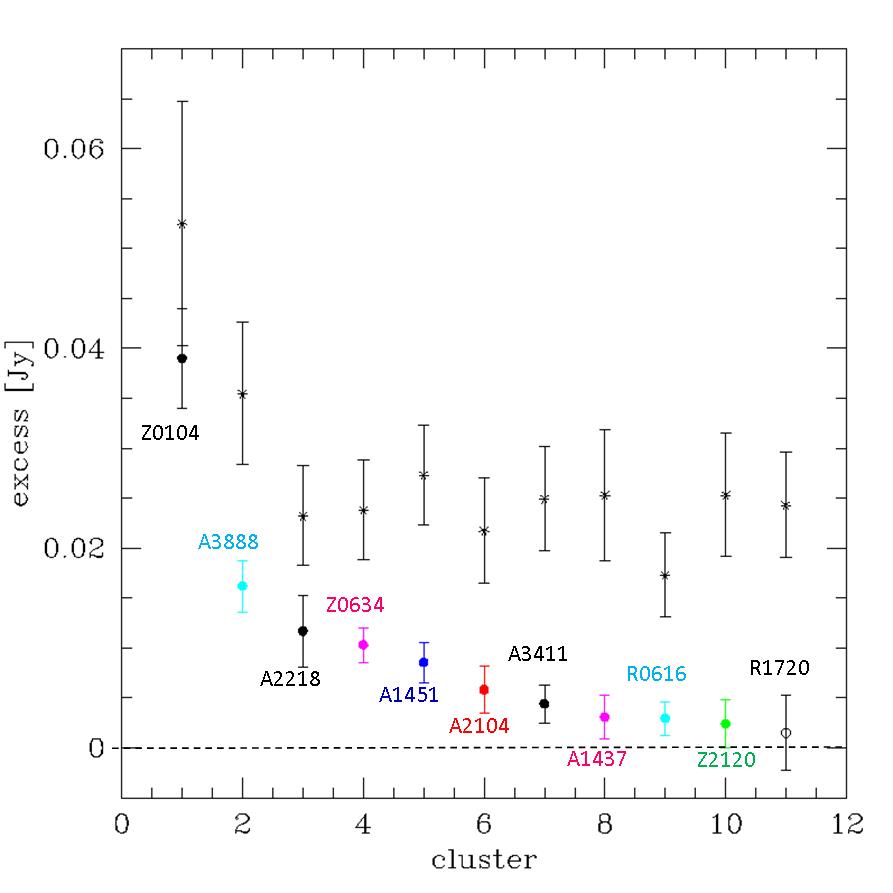}
\caption{\small{\textit{Left panel}: Diffuse flux density vs. control fields average flux density. Symbols: colored dots are clusters with reprocessed NVSS datasets; black filled dots are known RH clusters and the black open dot is the mini-halo. The red line is the 1:1 line. \textit{Right panel}: Excess of diffuse flux with respect to the average control fields flux densities. Black asterisks represent the expected radio power of the RH on the basis of the $P_{1.4}-M_{500}$ correlation (Cassano et al. 2013). }}\label{fig:pallini}
\end{figure*}

In Fig.~\ref{fig:pallini} (left panel) we compare the average value of the CFF with the RDF for each cluster, whereas the offset between the RDF and the average CFF are reported in Fig.~\ref{fig:pallini} (right panel).
Four clusters (A3888, Zwcl0634.1+4750, A1451 and A2104) and the clusters with already known RHs (A2218, Zwcl 0104.9+5350 and A3411), show an excess at $> 2 \sigma$ level. We consider this as the threshold level to identify clusters with the possible presence of a RH. We note that all the clusters show a positive offset (at least $\sim$few mJy) between the RDF and the CFF (Fig. ~\ref{fig:pallini}, right panel). This however is likely due to residual contamination from faint cluster radio galaxies that are below the NVSS detection limit, rather than to diffuse flux on cluster scales (\eg Farnsworth et al. 2013). For the sake of completeness, in Fig.~\ref{fig:pallini} (right panel, black asterisks) we also show the expected level of RH emission according to the $P_{1.4}-M_{500}$ correlation (Cassano et al. 2013, Eq. 4.10).   
We note that the mini-halo in RXJ1720.1+2638 does not result as an excess of diffuse emission. As explained above, this is due to the fact that a large fraction of the diffuse emission associated to the mini-halo is masked with the central bright radio galaxy.

Although with our procedure based on NVSS we can identify cases with suspect diffuse radio emission, we cannot confirm the presence of RHs in A3888, Zwcl0634.1+4750, A1451 and A2104. Deeper observations at low resolution (\eg VLA array C or D), in order to have a good sensitivity to the diffuse Mpc scale emission, and at high resolution (\eg VLA array A or B), to make an accurate subtraction of the individual sources from the \textit{u-v} data, are necessary.



\section{Cluster dynamical status } 
\label{sec:chandra}

In this Section we report on the analysis of the dynamical properties of clusters using \textit{Chandra} X-ray data. A high fraction  of clusters of the sample (50 out of 57) has \textit{Chandra} archival data. 24 of them already have dynamical information in the literature (Cassano et al. 2010, Cassano et al. 2013, see Tab.~\ref{tab:completesample}). We produced the X-ray images of the remaining 26 clusters (marked with $\surd$ in Tab.~\ref{tab:completesample}) in the 0.5-2 keV band using CIAO 4.5 (with calibration files from CALDB 4.5.8). We adopted an automatic algorithm for the identification of point sources which were then removed from images. Each image was then normalized for the exposure map of the observation, which provides the effective exposure time as a function of the sky position exposed on the CCD. In our analysis we did not correct for the background emission to treat the exposure-corrected images without introducing negative values in correspondence of pixels with zero counts. This procedure is sufficiently safe both because we are dealing with integrated quantities and since inside $R_{ap}$=500 kpc (see below) the images are largely dominated by the signal associated to the cluster emission. Typically, using the background estimates provided in Tab. 6.7 at http://asc.harvard.edu $<$http://asc.harvard.edu/$>$/proposer/POG/html/chap6.html, we
found that $\sim95\%$ (both in ACIS I and ACIS S) of the total counts in the $0.5-2$ Kev band are from the cluster.

Following Cassano et al. (2010, 2013), we studied the cluster substructures on the RH scale analysing the surface brightness inside an aperture radius $R_{ap}= 500$ kpc, since we are interested in the cluster dynamical properties on the scales where the energy is most likely dissipated.  We used three main methods: the power ratios (\eg Buote \& Tsai 1995; Jeltema et al. 2005; Ventimiglia et al. 2008; B\"{o}hringer et al. 2010), the emission centroid shift (\eg Mohr et al. 1993; Poole et al. 2006; O'Hara et al. 2006; Ventimiglia et al. 2008; Maughan et al. 2008; B\"{o}hringer et al. 2010), and the surface brightness concentration parameter (\eg Santos et al. 2008).

The power ratio represents the multipole decomposition of the two-dimensional mass distribution inside a circular aperture $R_{ap}$, centred on the cluster X-ray centroid. 
The power ratio can be defined as:
\begin{eqnarray}
P_0=[a_0~ln(R_{ap})]
\end{eqnarray}
where $a_0$ is the total intensity inside the aperture radius: $a_0=S(<R_{ap})$, $S(x)$ is the X-ray surface brightness, and
\begin{eqnarray}
P_m=\frac{1}{2m^2R_{ap}^{2m}}(a_m^2+b_m^2)
\end{eqnarray}
where the moments $a_m$ and $b_m$ are given by:
\begin{eqnarray}
a_m(R)=\int_{R'\leq R_{ap}}S(x')(R')cos(m\phi')d^2x'
\end{eqnarray}
and
\begin{eqnarray}
b_m(R)=\int_{R'\leq R_{ap}}S(x')(R')sin(m\phi')d^2x'
\end{eqnarray}
we will only make use of the $P_3/P_0$ parameter that is related to the presence of multiple peaks in the X-ray distribution providing a clear substructure measure (Buote 2001, B\"{o}hringer et al. 2010).

The centroid shift, $w$ is defined as the standard deviation of the projected separation between the peak and the centroid in unit of $R_{ap}$ and it is computed in a series of circular apertures centered on the cluster X-ray peak (\eg Poole et al. 2006):
\begin{eqnarray}
w=\left[\frac{1}{N-1}\sum(\Delta_i-\langle \Delta\rangle)^2\right]^{1/2}\times \frac{1}{R_{ap}}
\end{eqnarray}
$\Delta_i$ is the distance between the X-ray peak and the centroid of the \textit{ith} aperture.

Following Santos et al. 2008 we define the concentration parameter as the ratio between the peak and the ambient surface brightness:
\begin{eqnarray}
c=\frac{S(r<100~kpc)}{S(<500~kpc)}
\end{eqnarray}
The concentration parameter allows to distinguish clusters with compact core (not disrupted by recent mergers) from clusters with a spread distribution of the gas in the core.

Basically, high values of $P_3/P_0$ and $w$ indicate a dynamically disturbed system, while high values of $c$ stand for highly relaxed systems.

\section{Occurrence of Radio Halos}
\label{Sec:occurrence}
\begin{figure*}
   \centering
  \includegraphics[width=7cm]{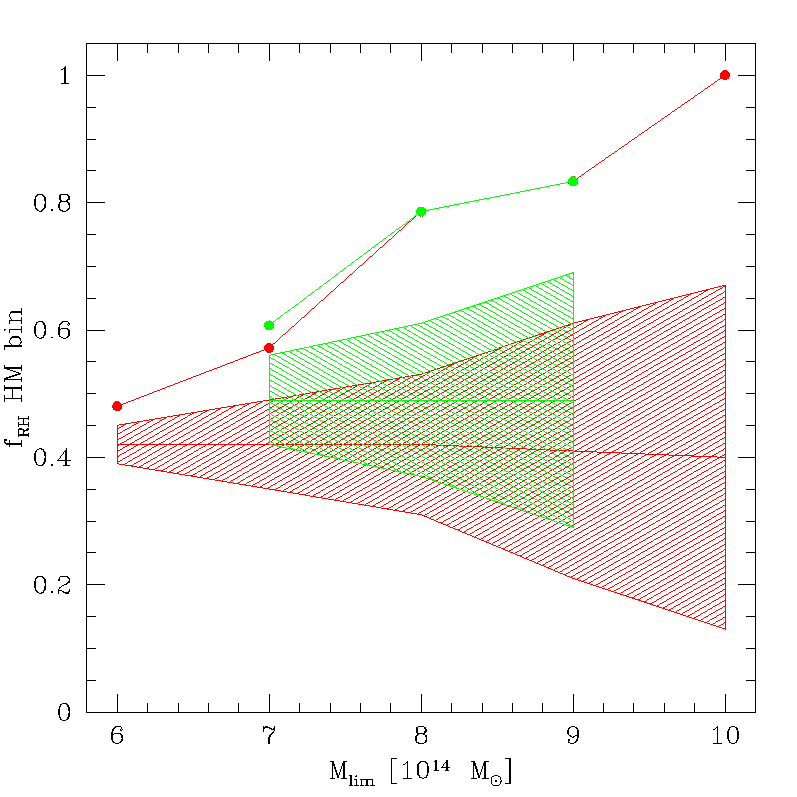}
  \includegraphics[width=7cm]{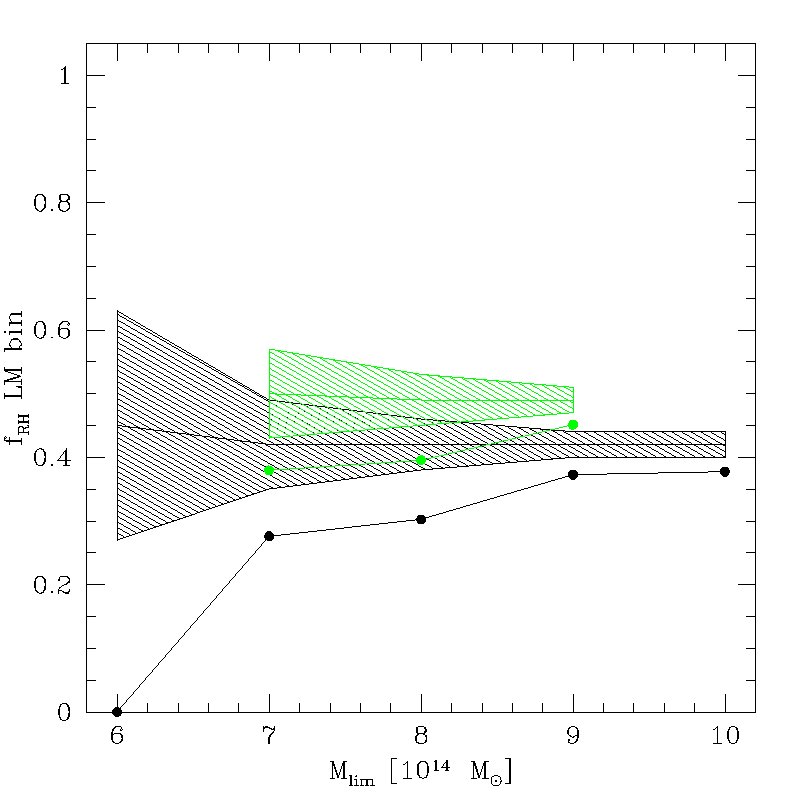}
\caption{\small {Observed fraction of RHs $f_{RH}$ (dots) compared to the value predicted by the Monte Carlo simulations (shadowed regions) in the HM bin (left panel) and in the LM bin (right panel) as a function of the limiting mass $M_{lim}$. In both panels the green dots and the green shadowed regions represent the case in which the 4 clusters with suspect diffuse emission are considered as RH clusters.} }\label{fig:fraction}   
   \end{figure*}

The aim of this Section is to derive the occurrence of RHs as a function of the mass of the hosting clusters. Among the sample of 57 clusters with radio information, 24 host RHs and 4 show residual emission in a Mpc-scale region that is a possible indication for the presence of a RH (Sect.~\ref{Sec:NVSS_analysis}). We split this sample into two mass bins and derived the fraction of clusters with RH, $f_{RH}$, in the low mass bin (LM, $M<M_{lim}$) and in the high mass bin (HM, $M>M_{lim}$) for different values of the limiting mass, $M_{lim}$ (as detailed below). In general we found that $f_{RH}$ is lower in the LM bins ($f_{RH}\approx 20-30$\%) while it is higher ($f_{RH}\approx60-80$\%) in the HM bins (Fig.~\ref{fig:fraction}).

\begin{figure}
   \centering
  \includegraphics[width=7cm]{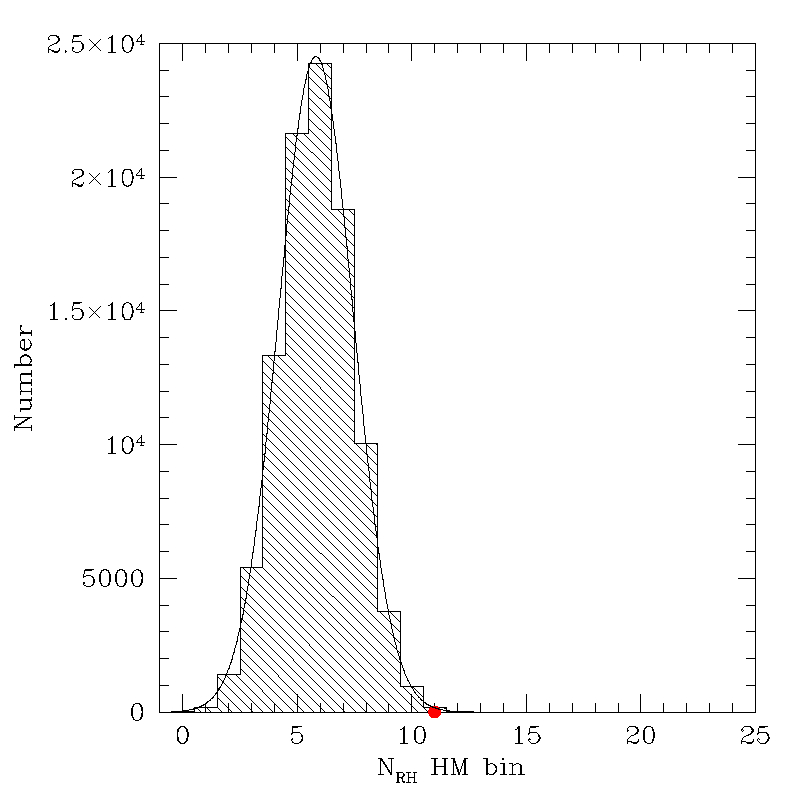}
   \caption{\small {Distribution of the number of RHs in the HM bin ($M>M_{lim}=8\times 10^{14} M_\odot$) after $10^5$ Monte Carlo trials. The red pont represents the observed number of RHs in the HM bin.}}
   \label{fig:montecarlo}
   \end{figure}

This difference is systematic and thus we attempted to identify the value of $M_{lim}$ that provides the most significant jump between low and high-mass clusters. We performed Monte Carlo simulations considering both the cases in which the four objects in the low-z sample with suspect diffuse emission are included (\textit{i}) as non RH clusters and (\textit{ii}) as RH clusters. Considering the case \textit{(i)}, we randomly assigned 24 RHs among the 57 clusters of the sample and obtained the distributions of RHs in the two mass bins (after $10^5$ trials), expected in the case that RHs were distributed independently of the cluster mass.
We consider 5 different values of the transition mass between the two bins, specifically $M_{lim}=(6, 7, 8, 9, 10)\times 10^{14}\,M_\odot$. An example of the expected distribution of the number of RH in the HM bin is shown in Fig. \ref{fig:montecarlo} for the case $M_{lim}=8\times10^{14}\,M_\odot$.
Each distribution can be well fitted by a gaussian function. The results of the Monte Carlo simulations are reported in Tab.~\ref{tab:sigma_gauss} for both the cases \textit{(i)} (upper panel) and \textit{(ii)} (lower panel). Specifically we report the number of clusters ($N_{clusters}$), the number of RHs ($N_{RH}$) and the fraction of clusters hosting RHs ($f_{RH}$) in the two mass bins for each value of $M_{lim}$. In Tab.~\ref{tab:sigma_gauss} we also report the significance of our result in unit of $\sigma$, $Z=(N_{RH}-\mu)/\sigma$ (where $\mu$ is the gaussian median value), and the most likely value of $f_{RH}$, $f_{\mu}$=$\mu/N_{cluster}$, in the HM bin (very similar results are obtained for the LM bin).



Fig.~\ref{fig:fraction} shows the observed fraction of RHs (dots) together with the results of the Monte Carlo simulations (shadowed regions) in the HM bin (left panel) and in the LM bin (right panel). We report the measured fraction of cluster with RHs and the results of the Monte Carlo analysis in both cases \textit{(i)} (red and black dots and shadowed regions) and \textit{(ii)} (green dots and shadowed regions). For a more clear visualization in Fig.~\ref{fig:fraction}, for the case \textit{(ii)} we only show the results obtained by assuming $M_{lim}= (7, 8, 9)\times10^{14}M_\odot$. Fig.~\ref{fig:fraction} shows that in the HM bin the observed $f_{RH}$ is always grater than that predicted by the Monte Carlo simulations, on the contrary in the LM bin the observed $f_{RH}$ is always lower than that predicted by the Monte Carlo analysis. This suggests the existence of a systematic drop of $f_{RH}$ in low mass systems.

In both cases \textit{(i)} and \textit{(ii)}, we found that the value of $M_{lim}$ that gives the most significant result and maximizes the drop of $f_{RH}$ between the two mass bins is $M_{lim}\approx 8\times 10^{14}M_\odot$, for which $f_{RH}\simeq30$\% (40\%) in the LM bin and $f_{RH}\simeq79$\% (79\%) in the HM bin in the case \textit{(i)} (in the case \textit{(ii)}). 
For $M_{lim}\approx 8\times 10^{14}M_\odot$ the observed $f_{RH}$ in the two mass bins differs from that obtained by the Monte Carlo analysis at $\sim3.2\sigma$ in the case \textit{(i)} and $\sim2.5\sigma$ in the case \textit{(ii)}. This means that the chance probability of the observed drop of $f_{RH}$ is $<7.4\times 10^{-4}$ \textit{(i)} and $<5.7\times 10^{-3}$ \textit{(ii)}.

 \noindent Based on this analysis we conclude that there is statistical evidence for a drop of the fraction of RHs in galaxy clusters at smaller masses. A similar conclusion was obtained using X-ray selected clusters (Cassano et al. 2008), however this is the first time that such indication is derived using a mass-selected sample. In Sect. 7 we discuss possible biases due to incompleteness in our current sample.

\begin{table*}
\begin{center}
\vspace{2mm}
\begin{tabular}{|c|c|c|c|c|c|c|c|c|c|}
\hline
case No. &$M_{lim}$&$N_{RH}$&  $N_{cluster}$& $N_{RH}$&$N_{cluster}$&  $f_{RH}$&$f_{RH}$& $f_{\mu}$&Z\\
&$(10^{14}\,M_\odot$)&(HM)  &(HM)&(LM)&(LM)  & (HM)&(LM)& (HM) &(HM)\\
\hline
&6& 24 &50& 0 &7  & $48\%$ & $0$\% & 42\%& 2.49  \\
\cline{2-10}
&7& 16 & 28 &8&29   & $57\%$ & $28\%$ & 42\%&2.29  \\
\cline{2-10}
\textit{(\textbf{i})}&8 & 11 &14 &13 &43& $79\%$ & $30\%$  & 42\%&3.17 \\
\cline{2-10}
&9 & 5 &6 &19 &51    &$83\%$ & $37\%$  &41\%& 2.07 \\
\cline{2-10}
&10 & 4 &4 &20 &53    &$100\%$ & $38\%$ & 40\%&2.21 \\
\hline
&6 & 27& 50 &1& 7& 54\% & 14\% & 49\% & 2.10\\
\cline{2-10}
&7 &17 & 28& 11 & 29 & 60\% & 38\% & 49\% & 1.76\\
\cline{2-10}
\textit{(\textbf{ii})}&8 & 11 & 14 & 17 & 43 & 79\% & 40\% & 49\% & 2.50\\
\cline{2-10}
&9 & 5 &  6 & 23 & 51 & 83\% & 45\% & 49\% & 1.72\\
\cline{2-10}
&10 & 4 & 4 & 24 & 53 & 100\% & 45\%& 48\% & 1.93\\
\hline

\end{tabular}
\vspace{2mm}
\caption {\small The four clusters with suspect diffuse radio emission are considered as non RH clusters in the upper panel (\textit{i}) and as RH clusters in the lower panel (\textit{ii}).}
\label{tab:sigma_gauss}
\end{center}
\end{table*}


\section{Radio Halo--cluster merger connection}
\label{Sec:RH-merger}
\begin{figure*}
  \centering
  \includegraphics[scale=0.4]{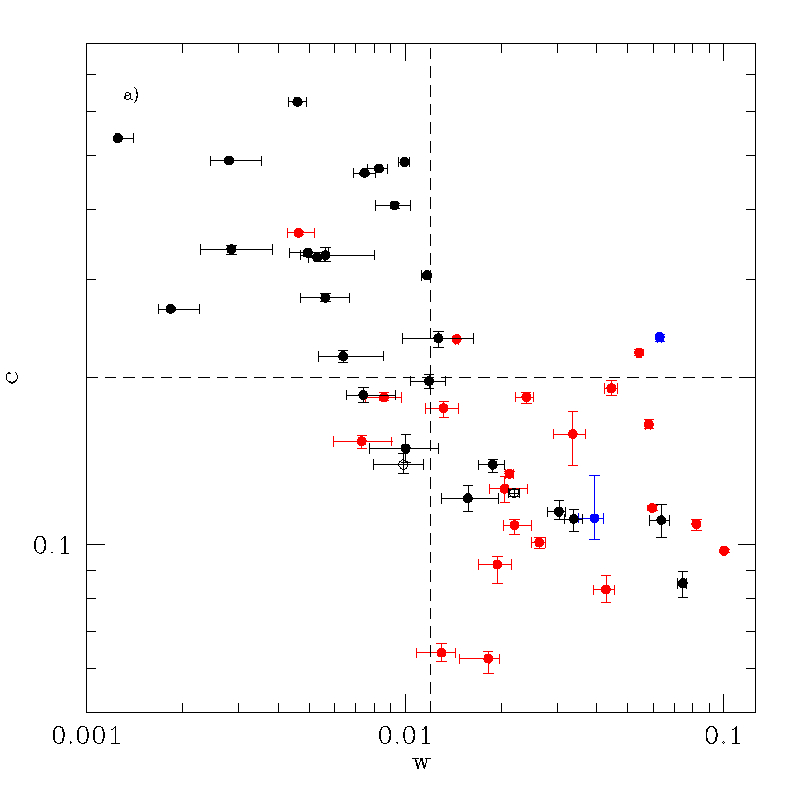}
  \includegraphics[scale=0.4]{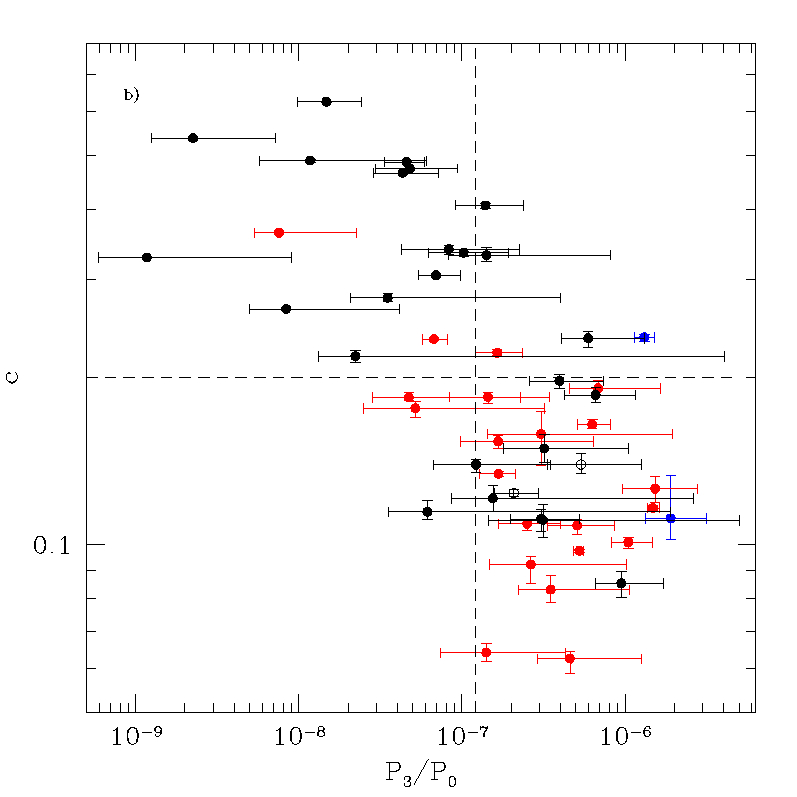}
  \includegraphics[scale=0.4]{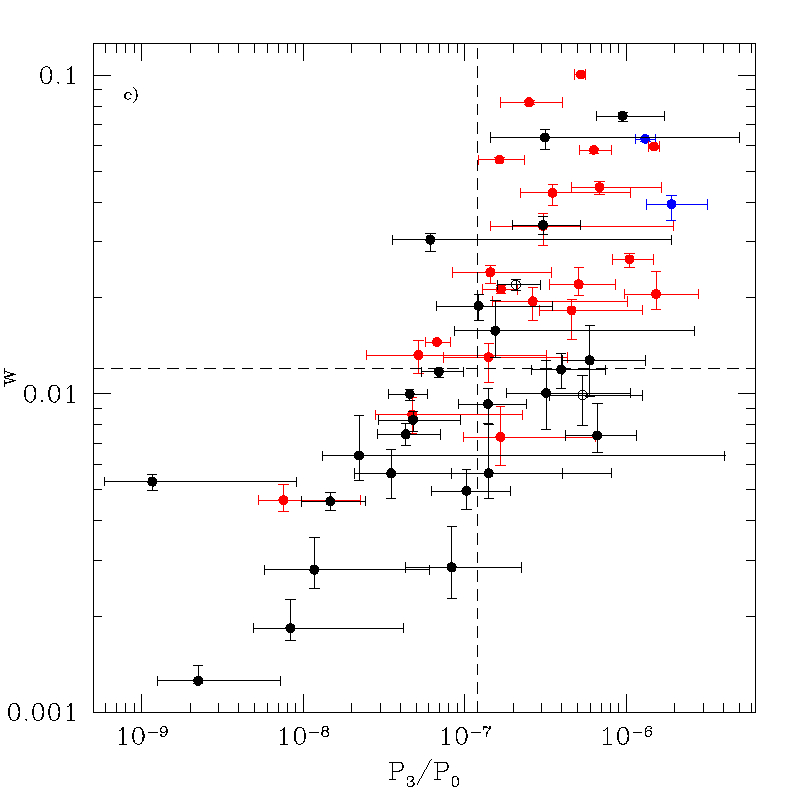}
  \caption{\small {(a) $c-w$, (b) $c-P_{3}/P_{0}$, (c) $w-P_{3}/P_{0}$ diagrams. Vertical and horizontal dashed lines: $c=0.2$, $w=0.012$ and $P_{3}/P_{0}=1.2\times10^{-7}$. Red, black and blue dots represent clusters with RH, clusters without RH and clusters hosting relics (without RHs), respectively. Black open dots are clusters with suspect diffuse emission from the NVSS.}}\label{fig:morp_tot}

\end{figure*}

In this Section we investigate the connection between the presence/absence of RHs in clusters and the cluster dynamical status merger/relaxed. Following Cassano et al. (2010),  in Fig.~\ref{fig:morp_tot} we report the cluster morphological parameters, derived in Sect.~\ref{sec:chandra} for the 50 clusters of the sample with available \textit{Chandra} data, in three diagrams: $c-w,~ c-P_3/P_0,~ w-P_3/P_0$. Vertical and horizontal dashed lines are taken from Cassano et al. (2010), these lines represented the median value of each parameter, and were used to separate merging ($w>0.012$, $c<0.2$ and $P_3/P_0>1.2\times10^{-7}$) and relaxed ($w<0.012$, $c>0.2$ and $P_3/P_0<1.2\times10^{-7}$) clusters. Here we use these lines as a reference to compare our measurements with previous published results\footnote{We note that Cassano et al. (2010) derived these lines on a smaller redshift range ($0.2-0.35$), however there is no clear indication about an evolution of $P_3/P_0$ and $w$ with $z$ (Wei\ss mann et al. 2013b and references therein)}. Fig. \ref{fig:morp_tot} shows that RH clusters (red dots) can be separated from clusters without RH (black dots) in the morphological diagrams: RHs are associated with dynamically disturbed clusters, while the greatest majority of clusters without Mpc-scale diffuse radio emission are relaxed objects. About 80\% of the clusters in the HM bin of our sample are mergers, and this explains why RHs are fairly common in this bin (Sect.~\ref{Sec:occurrence}). The only RH cluster that always falls in the region of relaxed clusters is A1689, however this cluster is undergoing a merger event at a very small angle with the line of sight (\eg Andersson \& Madejski, 2004), therefore its morphological parameters are likely biased due to projection effects. We note that also clusters with relics and without RHs (blue dots) are located in the regions of dynamically disturbed systems, in line with literature observations (\eg de Gasperin et al. 2014). 

We also note that at least 10 merging clusters of our sample do not host RHs. The existence of massive and merging systems without RHs is well known (Cassano et al. 2010, 2013, Russell et al. 2011). If RHs are due to turbulence acceleration of relativistic electrons during cluster mergers they should have a typical lifetime of $\sim$ Gyr (see Brunetti et al. 2009), which is of the same order of the merger timescale. However the generation (and cascading from large to smaller scales) of turbulence and its dissipation take some time, corresponding to a ``switch-on'' and ``switch-off'' phases that span a substantial fraction of a Gyr. This produces a partial ``decoupling'' between X-rays and radio properties, as during these phases RHs would appear underluminous/absent whereas the hosting cluster would appear disturbed in the X-rays (\eg Donnert et al. 2013).

\noindent

An additional possibility is that some of the dynamically disturbed systems host RHs with very steep spectrum, that are not easily seen at our observing frequencies (Cassano et al. 2006, Brunetti et al. 2008). Indeed the great majority of merging clusters without RHs belong to the LM bin, which might support the idea that in these cases (or some of them) the energy provided by the merger is not sufficient to generate RHs emitting at the observing frequencies.
In fact, this second possibility is expected to contribute to the drop of the fraction of RHs in less massive systems (Cassano et al. 2010, 2012), as currently observed in our sample (Sect.~\ref{Sec:occurrence}).


\section{Sample completeness}

\label{sec:completeness}

As explained in Sec. \ref{sec:sample}, in the calculation of the occurrence of RHs we included only clusters with available radio information about the presence/absence of RHs: these are 21/21 clusters in the low-z sample ($z<0.2$) and 36/54 in the high-z ($z>0.2$) sample. Considering that the PSZ sample gives a completeness of $\sim 90\%$ for the low-z sample and $80\%$ for the high-z sample, we can estimate a completeness of our sample (which takes into account both the completeness in mass and in the radio information) of $\sim63\%$\footnote{This is estimated as $\frac{21+36}{(21/0.9)+(54/0.8)}\sim 63\%$}.
18 clusters in the high redshift range ($z>0.2$) still lack radio information: 17 with $M<8\times 10^{14}M_\odot$ and one with $M>8\times 10^{14}M_\odot$. 

In this Section we evaluate how much the omission of these clusters can affect our results. 

We consider the total sample of 75 clusters and assume three extreme cases:
\begin{itemize}
\item[\textit{a)}] all the missing clusters  with $M<8\times 10^{14}M_\odot$ are clusters without RH and the only one with $M>8\times 10^{14}M_\odot$ is a RH cluster;
\item[\textit{b)}] all the missing clusters with $M<8\times 10^{14}M_\odot$ host a RH, and the one with $M>8\times 10^{14}M_\odot$ is a non RH cluster.
\item[\textit{c)}] the fraction of RHs in the 18 missing clusters is independent of the cluster mass.
\end{itemize}

Cases \textit{(a)} and \textit{(b)} are simply adopted to obtain the maximum \textit{(a)} and the minimum \textit{(b)} drop of the RH fraction with mass that can be expected starting from current data.
We stress however that the case \textit{(b)} is particularly unlikely since it implies that the occurrence of RHs is stronger in less massive systems, which is not justified by any observational results achieved so far (\eg Cassano et al 2008). 

We examined the scenarios listed above in both the cases where the 4 low-z clusters with suspect diffuse emission are considered as \textit{(i) }non RH clusters and as \textit{(ii)} RH clusters.   

In case \textit{(a)} we add 17 non RH clusters to the LM bin and one RH cluster to the HM bin, thus the fraction of clusters with RHs in the LM bin becomes $f_{RH}=13/60=22$\% ($17/60=28$\%), while in the HM bin $f_{RH}=12/15=80$\% ($12/15=80$\%) in case \textit{(i)} (in case \textit{(ii)}). Adopting the Monte Carlo approach, described in Sect.~\ref{Sec:occurrence}, we find that this corresponds to a 4.2 $\sigma$ (3.7$\sigma$) result in case \textit{(i)} (in case \textit{(ii)}).\\
In case \textit{(b)} we add 17 RH clusters to the LM bin and one non RH cluster to the HM bin, so that $f_{RH}(LM)=30/60=50$\% ($34/60=57$\%) and $f_{RH}(HM)=11/15=73$\% ($11/15=73$\%), with a 1.6 $\sigma$ (1.2 $\sigma$) significance level in case \textit{(i)} (in case \textit{(ii)}).

Finally, in case \textit{(c)}, we assumed that the fraction of RHs in the 18 missing clusters is the same we measured in the sample of 57 clusters analysed in the present paper: $\sim 42\%$ \textit{(i)}, $\sim 49\%$ \textit{(ii)}, independently of the cluster mass. The fraction of cluster with RHs in the LM bin would be $f_{RH}(LM)=20/60=33$\% ($25/60=42$\%) and in the HM bin $f_{RH}(HM)=11/15=73$\% ($11/15=73$\%), corresponding to a 2.8 $\sigma$ (2.2$\sigma$) result in case \textit{(i)} (in case \textit{(ii)}).

Based on our analysis we conclude that the evidence for a drop of the fraction of clusters with RH at smaller masses is  tempting and cannot be completely driven by possible biases deriving from the incompleteness in mass of the (radio) sample. Namely, even in the very unlikely and extreme case (\textit{b}), a hint of difference in the occurrence of RH still remains between the high-mass and low-mass systems in our sample.

\section{Summary \& Conclusions}
The study of the statistical properties of RHs in galaxy clusters has became increasingly important in the last decade: it is a powerful tool to test the theoretical models for their origin and to unveil the connection between RHs and cluster formation. In their simplest form, homogeneous re-acceleration models predict that RHs should be found in massive and merging objects, whereas the fraction of clusters with RHs, $f_{RH}$, should drop towards smaller merging systems and RHs should be absent in relaxed clusters. In order to test these expectations large mass-selected samples of galaxy clusters are necessary. SZ-cluster surveys, \textit{i.e.} the Planck-SZ survey (PSZ, Planck Collaboration 2014) have recently enabled the construction of cluster samples that are almost mass-selected, thanks to the tight correlation between the SZ signal and the cluster mass (Motl et al. 2005; Nagai 2006). Recent studies, based on the EGRHS (Venturi et al. 2007,2008, Kale et al. 2013,2015) and the PSZ catalogue, have shown the presence of a bimodal split between clusters with and without RH also in the radio-SZ diagrams for clusters with $Y_{500}>6\times 10^{-5}$ Mpc$^2$ (Cassano et al. 2013). However, the mass completeness of the sample used by Cassano et al. (2013) is ~50\% and does not allow to probe the existence of a drop in $f_{RH}$ towards small clusters. 

Here we have presented a step toward an unbiased analysis of the occurrence of RHs, as a function of the cluster mass, in a mass-selected sample of galaxy clusters. We built a sample of 75 clusters with $M\gtrsim 6\times 10^{14}M_\odot$ in the redshift range $0.08<z<0.33$ selected from the Planck SZ catalogue. Among these clusters 57 have available radio information, for 21/21 clusters in the redshift range $0.08-0.2$ we used NVSS and literature information, whereas 36/54 clusters at $z=0.2-0.33$ have data from the EGRHS (plus literature information). Our study is based on these 57 clusters. The completeness in mass of this sample is $\sim63$\%, larger than that available in previous studies (e.g. Cassano et al. 2013). We also used the available \textit{Chandra} X-ray data, for 50 out of 57 clusters, to derive information on the cluster dynamical status. \\
The presence/absence of RHs has been determined by using literature information for all the high-z clusters ($z>0.2$, Venturi et al. 2007,2008; Kale et al. 2013,2015 and references therein), and for the majority (14) of low-z clusters ($z<0.2$). We reprocessed and analysed NVSS data of the remaining 7 low-z clusters that lack literature information and conclude for possible diffuse emission in 4 cases (Sect.~\ref{Sec:NVSS_analysis}).
We split our sample into two mass bins, the low mass bin (LM, $M<M_{lim}$) and the high mass bin (HM, $M>M_{lim}$) and derived the fraction of clusters with RHs in the two mass bins for different values of $M_{lim}$, finding that $f_{RH}$ is $\approx 60-80$\% in the HM bin and $\approx 20-30$\% in the LM one. 
By means of Monte Carlo simulations we obtained the distributions of RHs in the two mass bins (after $10^5$ trials), expected in the case that RHs were distributed independently of the cluster mass.
We found that for $M_{lim}\approx 8\times 10^{14}M_\odot$ the observed $f_{RH}$ in the two mass bins differs from that obtained by the Monte Carlo analysis with a significance that ranges between 2.5 $\sigma$ and 3.2$\sigma$, which means that the probability to obtain by chance the observed drop of $f_{RH}$ is $<5.7\times 10^{-3}$ or even lower (see Sect~\ref{Sec:occurrence}). This highlights the statistical significance of our results and suggests that the increase of the occurrence of RHs with the cluster mass is likely to be real, rather than by chance.

\begin{figure}
  \centering
  \includegraphics[scale=0.4]{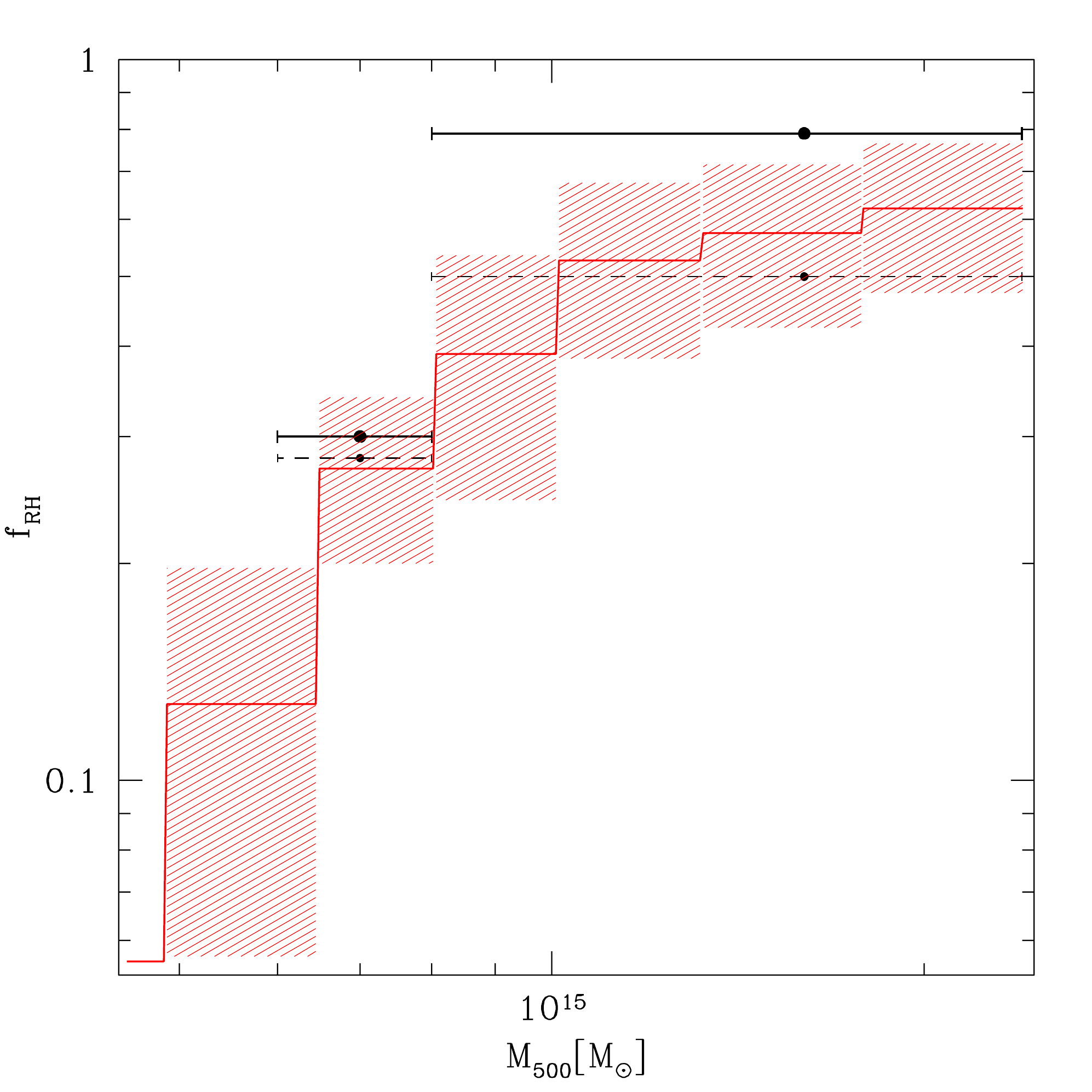}
  \caption{\small {Expected fraction of clusters with RHs with steepening frequency (Cassano et al., 2010) $\nu_s>600$ MHz in the redshift range $0.08<z<0.33$ (red line and shadowed region). Calculations have been performed for the following choice of model parameters: $b=1.5$, $\langle B\rangle=1.9~\mu$G (where $B=\langle B\rangle\times (M/\langle M\rangle)^b)$ and $\eta_t=0.2$ (see Cassano et al. 2012 and referencees therein). The observed fraction of clusters with RHs in the two mass bins is also shown (black points with horizontal errorbars). The balck points with dashed errorbars show $f_{RH}$ in the case we exclude USSRH (and candidate USSRH) from the observed sample.}}\label{fig:model}
\end{figure}

The possibility of a drop of the fraction of clusters hosting RH for less massive systems is particularly intriguing. Indeed this is naturally and uniquely expected in the framework of turbulent re-acceleration models (\eg Cassano \& Brunetti 2005) that provide a popular picture for the formation of giant RHs in galaxy clusters.
A solid comparison between models and our observations is still premature due to the incompleteness of the observed sample (Sect.~\ref{sec:completeness}). Still, with this caveat in mind, in Fig.~\ref{fig:model} we compare our measurements of the occurrence of RHs in the two mass bins (black solid line) with the formation probability of RHs derived from the turbulent re-acceleration model in its simplest form (red line). Specifically, following Cassano \& Brunetti (2005), we adopted the semi-analytic Press \& Schechter theory (PS, Press \& Schechter, 1974) to generate merger-trees and follow the hierarchical evolution of galaxy clusters through merger events. We assumed that a fraction, $\eta_t$ of the $PdV$ work done by the infalling subclusters during mergers is channelled into magnetosonic waves that accelerate relativistic electrons, which in turn emit synchrotron radiation. We calculated the theoretical evolution of $f_{RH}$ with the cluster mass in the redshift range $z=0.08-0.33$ for given values of the model parameters (see caption of Fig.~\ref{fig:model} for details). Uncertainties on the predicted formation probability are estimated by running MC extractions from the pool of theoretical merger trees and accounting for the statistical variations that are induced by the limited size of the two observed subsamples defined in Tab.~\ref{tab:sigma_gauss} (using $M_{lim}=8\times 10^{14}M_\odot$). Despite the crude approximations adopted in these models, there is an overall agreement between the observed and predicted behaviour of $f_{RH}$ with the cluster mass. 
The model slightly underestimates $f_{RH}$ in the high mass bin. This may be due to two main reasons: 1) the use of the PS formalism, which is well know to underestimate the merging rate, and hence the number density, of very massive systems; 2) the fact that the model predictions do not include RHs with very steep radio spectra, \ie those with steepening frequency $\nu_s\ltsim 600$ MHz. As an example in Fig.~\ref{fig:model} (black dashed lines) we report the effect on the observed statistics induced by removing from the sample USSRHs (marked with ``\textit{US}'' in Tab.~\ref{tab:completesample}) and candidate USSRHs (marked with ``\textit{c}'' in Tab.~\ref{tab:completesample}), for which we don't know the detailed spectral shape.

With the procedure described in Sec. \ref{sec:chandra} we analysed the \textit{Chandra} X-ray data of 26 clusters and we derived the morphological parameters (the centroid shift, $w$, the power ratio, $P_3/P_0$ and the concentration parameter, $c$), which are powerful diagnostic of the cluster dynamical status. We combined them with previously published results (Cassano et al. 2010, 2013) and we confirmed that RHs are hosted by merging clusters, while the majority of non-RH clusters are relaxed, thus highlighting the key role that merger events play in the origin of RHs. We note the presence of few merging clusters without RHs. This observational fact adds constraints for the origin and evolution of RHs that have been briefly discussed in Sect.~\ref{Sec:RH-merger} in the context of current models. 

The calculation of the occurrence of RHs has been performed only for clusters with radio information about the presence of diffuse radio emission in form of RH. 18 clusters are not included in our analysis because they still lack radio information. In Sec.~\ref{sec:completeness} we tested the possible effects of the sample incompleteness on our results assuming three different situations for the derivation of the final $f_{RH}$ (see Sect.~\ref{sec:completeness} for details).
We found that even in the most unfavourable case, that however is very unlikely (case (\textit{b}) in Sect.~\ref{sec:completeness}), a drop of the fraction of RHs at smaller masses would still remain.

This is the first step of this study; observations of the missing  clusters with the GMRT and the VLA are already in progress and will allow the conclusive measure of the occurrence of RHs in a mass-selected sample of galaxy clusters.


\begin{acknowledgements}
The National Radio Astronomy Observatory is a facility of the National Science Foundation operated under cooperative agreement by Associated Universities, Inc. We thank the staff of the GMRT that made these observations possible. GMRT is run by the National Centre for Radio Astrophysics of the Tata Institute of Fundamental Research. The scientific results reported in this article are based in part on data obtained from the \textit{Chandra} Data Archive. This research has made use of the NASA/IPAC Extragalactic Database (NED) which is operated by the Jet Propulsion Laboratory, California Institute of Technology, under contract with the National Aeronautics and Space Administration. GB acknowledges support from the Alexander von Humboldt Foundation.
\end{acknowledgements}


\end{document}